\documentclass[sigconf]{acmart}
\usepackage{graphicx}
\usepackage{subcaption}
\usepackage{CJKutf8}
\usepackage{longtable}
\usepackage{tabularx} 
\usepackage{booktabs} 
\usepackage{xcolor}   
\usepackage{seqsplit}
\definecolor{botGenColor}{RGB}{0, 85, 128}    
\definecolor{botPresetColor}{RGB}{153, 0, 76} 
\definecolor{intervieweeColor}{RGB}{30, 30, 30} 

\newcommand{\interviewee}[1]{%
    \textcolor{intervieweeColor}{#1}%
}

\usepackage{tabularx}
\usepackage{booktabs}
\usepackage{xcolor}
\usepackage{balance}

\definecolor{botGenColor}{RGB}{124, 85, 128}

\definecolor{botPresetColor}{RGB}{100, 0, 76}

\definecolor{botFollowUpColor}{RGB}{204, 85, 0}

\definecolor{intervieweeColor}{RGB}{30, 30, 30}

\definecolor{triggerColor}{RGB}{0, 128, 128}

\newcommand{\botcontext}[1]{\textcolor{botGenColor}{#1}}

\newcommand{\botpreset}[1]{\textbf{\textcolor{botPresetColor}{#1}}}

\newcommand{\botfollowup}[1]{\textbf{\textcolor{botFollowUpColor}{#1}}}

\newcommand{\trigger}[1]{\textbf{\textcolor{triggerColor}{#1}}}
\AtBeginDocument{%
  }

\setcopyright{acmlicensed}
\copyrightyear{2026}
\acmYear{2026}
\setcopyright{cc}
\setcctype{by}
\acmConference[HCOMP 2026]{2026 ACM Conference on Human-AI Complementarity and Alignment}{September 27--30, 2026}{Alexandria, VA, USA}
\acmBooktitle{2026 ACM Conference on Human-AI Complementarity and Alignment (HCOMP 2026), September 27--30, 2026, Alexandria, VA, USA}
\acmDOI{10.1145/3834580.3838754}
\acmISBN{979-8-4007-2894-5/2026/09}

\begin{document}

\title{When the Interviewer Is a Bot: \\Behavior, Breakdowns, and Trust in MLLM-Led Interviews}

\author{He Zhang}
\orcid{0000-0002-8169-1653}
\email{hpz5211@psu.edu}
\affiliation{%
\department{College of Information Sciences and Technology}
  \institution{The Pennsylvania State University}
  \city{University Park}
  \state{Pennsylvania}
  \country{USA}
  \postcode{16802}
}

\author{Kambinachi Chukwuma}
\orcid{0009-0007-3802-9828}
\email{kfc5673@psu.edu}
\affiliation{%
\department{College of Engineering}
  \institution{The Pennsylvania State University}
  \city{University Park}
  \state{Pennsylvania}
  \country{USA}
  \postcode{16802}
}

\author{ChanMin Kim}
\orcid{0000-0001-9383-8846}
\email{cmk604@psu.edu}
\affiliation{%
\department{College of Education}
  \institution{ Pennsylvania State University}
  \city{University Park}
  \state{Pennsylvania}
  \country{USA}
  \postcode{16801}
}
\author{John M. Carroll}
\orcid{0000-0001-5189-337X}
\email{jmc56@psu.edu}
\affiliation{%
\department{College of Information Sciences and Technology}
  \institution{Pennsylvania State University}
  \city{University Park}
  \state{Pennsylvania}
  \country{USA}
  \postcode{16802}
}

\renewcommand{\shortauthors}{He Zhang et al.}

\begin{abstract}

Semi-structured interviews are a cornerstone of qualitative research but remain labor-intensive. We report an empirical study of what actually happens when the interviewer is an off-the-shelf real-time multimodal LLM (MLLM). We built InterviewBot, a voice-based interviewing system that wraps a real-time MLLM with a researcher-authored outline, and deployed it not as a novel architecture but as a research instrument for observing default MLLM interviewing behavior. In a practice study ($N=15$), participants completed a bot-led semi-structured interview and then a human-led reflection session about that experience. We contribute (i) a turn-level behavioral analysis of an MLLM interviewer ($N_{\text{turns}}=428$) showing that it is acknowledgment-heavy but probe-light (deepening probes account for 4.9\% of all turns), and that 28.7\% of question-bearing turns pack multiple questions into one turn despite an explicit one-question-at-a-time instruction; (ii) an inductive catalogue of four data-collection breakdowns (information loss, premature termination, latency, and interruption) observed in a deployed rather than simulated system; and (iii) three social dynamics from participants' reflections: \emph{disclosure calibration}, where reduced social pressure coincided with shallower elaboration; \emph{institutional legitimacy}, where trust tracked perceived stakes and what delegation to AI signaled about the organizer rather than conversational competence; and \emph{conversational grounding}, where content-grounded paraphrase, not generic social filler, was what participants read as listening. We conclude with design implications for depth control, transparent handoffs, and non-templated listening mechanisms in human-centered interview automation.
\end{abstract}



\begin{teaserfigure}
    \centering
    \includegraphics[width=0.8\linewidth]{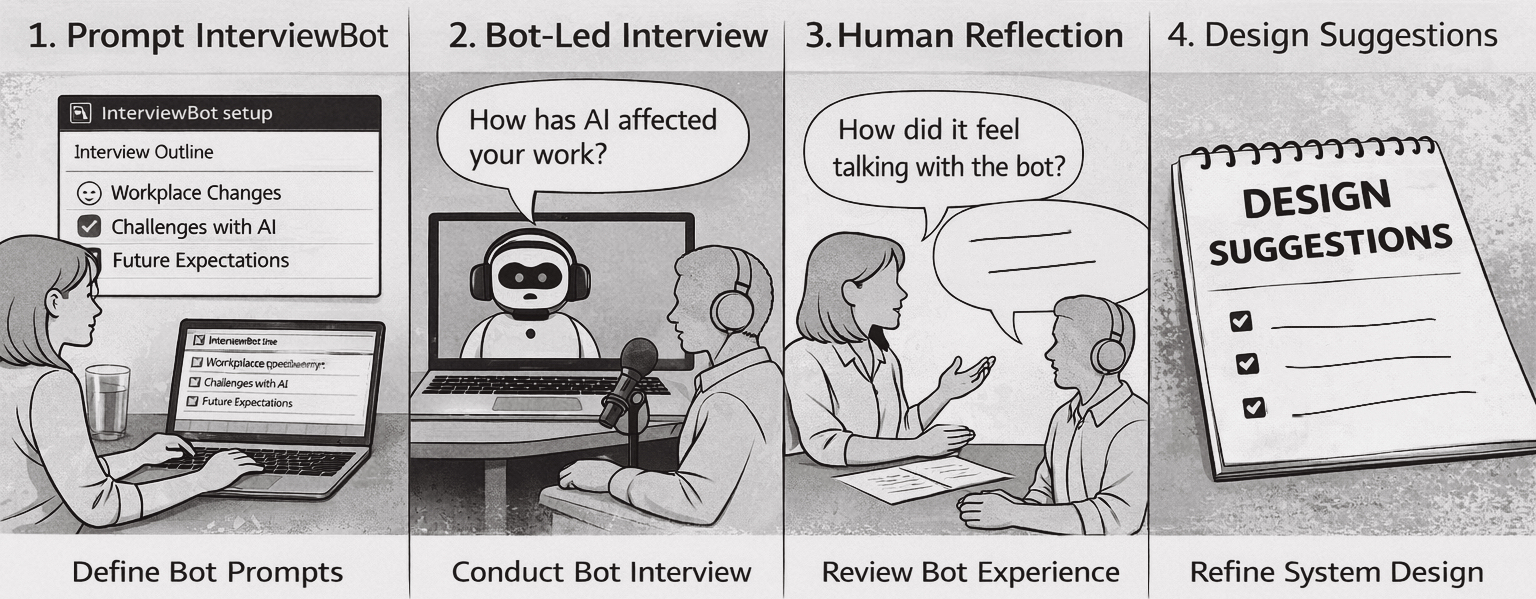}
    \caption{An example of study workflow with InterviewBot: (1) the researcher configures InterviewBot by entering interview prompts and an outline; (2) InterviewBot conducts an outline-guided, real-time interview with the participant; (3) the researcher conducts a post-session reflection interview to capture the participant's experience with the bot-led interview; and (4) insights from transcripts and reflections are synthesized into design implications to refine InterviewBot and inform human-centered interview automation.}
    \vspace{-0.1cm}
    \label{fig:teaser}
\end{teaserfigure}
%

\maketitle

\section{Introduction}\label{sec:intro}

Qualitative interviews are one of the most widespread data collection methods, combining a structured approach with the flexibility of open-ended unstructured exploration~\cite{roulston2010}. However, they are time-consuming and labor-intensive, and are often influenced by interviewer-related factors~\cite{flick2006, marshall2014, adams2015}. While early automated systems relied on rigid templates~\cite{10.1145/3637320, 10.1145/3382507.3418839,10.1145/2493432.2493502}, recent research on Large Language Model (LLM)-based systems has demonstrated that AI can generate interview questions and follow-up prompts that are significantly more human-like, linguistically diverse, and contextually relevant~\cite{ZHANG2025100144,10.1145/3686215.3688377,spangher-etal-2025-newsinterview, liu2025aiinterviews}. In practice, however, researchers require tools that preserve human control, provide transparency, and integrate smoothly into existing research workflows, rather than systems that aim to fully replace human interviewers. These considerations stem not only because interviews are social encounters shaped by trust, power relations, respect, and accountability rather than simple information-extraction procedures~\cite{kvale2006dominance,karnieli2009power}, but also from a range of concerns associated with AI use, including but not limited to hallucinations, ethical risks, and potential impacts on interview quality~\cite{liu2023speech,10.1145/3581641.3584051,biswas2024hi}.

To understand how automated interviewers behave in practice, and how participants experience being interviewed by one, we present \textbf{InterviewBot}, a real-time voice-based system that conducts semi-structured interviews from a researcher-authored outline. Through a practice study with $N{=}15$ participants, we examined the system's turn-level conduct and how users experienced a bot-led interview.

\paragraph{Contributions and positioning.}
We want to be explicit about what this paper does and does not claim. InterviewBot is not offered as a technical contribution: it is a deliberately thin wrapper around a commercial real-time MLLM, and its value here is as an \emph{instrument} that lets us observe what a researcher would obtain from such a model out of the box. Our contribution is empirical and twofold. Behaviorally, we characterize the dialogue-act profile and failure modes of a deployed (rather than Wizard-of-Oz) MLLM interviewer at the level of individual turns. Socially, we characterize how being interviewed by a bot reorganizes participants' sense of obligation, trust, and what the interview \emph{means}---dynamics that, unlike raw conversational fluency, are unlikely to be resolved by the next model release. We treat the specific proportions reported here as a snapshot of one model version rather than a stable property of MLLMs. Section~\ref{sec:interactionanalysis} establishes what the bot did; Section~\ref{sec:findings}, which we regard as the paper's primary contribution, establishes what it meant to be interviewed by it.

\section{Background of AI-supported Interview}
Before the rise of generative models such as LLMs, computational support for interviewing primarily drew on advances in NLP, including Transformers, BERT, BART, and T5, which enabled early work on automatic question generation, rule-based question design, and scripted dialogue systems~\cite{jain2021survey,lewis-etal-2020-bart,JMLR:v21:20-074,10.5555/1857999.1858085,debnath2020designing,sb2020automatic,10.1145/3637320,10.1145/505282.505285}. In parallel, HCI research explored embodied agents, virtual interviewers, chatbot-assisted surveys, and conversational probes as ways to scaffold or mediate interview practice~\cite{10.1145/2493432.2493502,10.1145/3411764.3445116,10.1145/3290605.3300705,10.1145/3613904.3642707,10.1145/3232077,10.1145/3313831.3376131,xiao2020tell,10.1145/3411764.3445569}. More recent AI-powered systems introduced real-time follow-up generation and theory-driven probing, demonstrating potential for greater efficiency and consistency, while also raising concerns around interviewer control, rapport, and ethical responsibility~\cite{10.1145/3382507.3418839,biswas2024hi,liu2023speech,10.1145/3581641.3584051}. However, across these stages, existing systems have generally struggled to handle the dynamic and context-sensitive nature of real interviews, and limitations in user experience remain evident even in widely deployed conversational systems such as telephone customer service platforms~\cite{qin2025customerservicerepresentativesperception,adam2021ai}.


Recent advances in LLMs have expanded the role of AI in qualitative research through techniques such as structured prompting, chain-of-thought prompting, retrieval-augmented generation, and RLHF, which improve contextual grounding and response reliability~\cite{bai2022traininghelpfulharmlessassistant,liu2024llms,ZHANG2025100144,10762977,gao2024retrievalaugmentedgenerationlargelanguage}. Researchers have begun exploring LLMs in interview-like settings, where they can generate follow-up questions and clarifications, yet important concerns remain regarding interview flow, interviewer authority, trust, timing, and accountability in hybrid human-AI interviewing~\cite{liu2025aiinterviews,10.1145/3686215.3688377,spangher-etal-2025-newsinterview}. This line of work has begun to consolidate into practical guidance for researchers adopting LLM support across the qualitative workflow~\cite{10.1145/3742414.3794947}. It has also surfaced concerns that originate with interviewers rather than with participants: in an LLM-in-the-loop Wizard-of-Oz study in which a co-interviewer relayed AI-generated follow-up questions during live sessions, interviewers raised worries about harmful or discriminatory phrasing, about divided attention and missed nonverbal cues eroding interviewees' sense of respect, about unequal participation, about who is accountable when harm occurs, and about privacy when an AI system listens and transcribes~\cite{10.1145/3805029.3818282}. Our study approaches the same setting from the other side of the exchange, asking how participants experience being interviewed by such a system once it is deployed without a human relay.

Importantly, this literature has largely been evaluated on the \emph{output} of automated interviewing: chatbot-administered surveys elicit more informative open-ended answers than standard web forms~\cite{xiao2020tell}, and recent LLM-based interviewers elicit responses comparable to human-conducted interviews at substantially lower cost~\cite{chopra2023conducting}. AI-supported interviewing has also recently gained visibility at scale; Anthropic Interviewer, for example, demonstrated the feasibility of multilingual, large-scale AI-mediated interviewing while converting open-ended responses into more structured and analyzable outputs~\cite{huang2026interviewer}. What remains less examined is the interviewer's \emph{conduct}---the moment-to-moment dialogue acts through which an interview is actually accomplished---and the social meaning participants assign to being interviewed this way. Those are the two gaps this paper addresses.

\section{InterviewBot as a Research Instrument}\label{sec:system}

InterviewBot is an MLLM-powered system that facilitates semi-structured interviews by balancing predefined outlines with dynamically generated, vocalized follow-up questions. It supports natural spoken interaction for participants, while providing researchers with text-based configuration, real-time session control, and live transcription.

In this work, the ''Bot'' operates strictly as a conversational, voice-based software agent. We deliberately avoided physical or anthropomorphic avatars to minimize unintended social pressure and attribution effects. This lack of embodiment ensures our study isolates how participants respond to the AI-mediated interview process itself, rather than reacting to the simulated personality of a humanoid agent.

\subsection{Interface and Configuration}\label{sec:interface}

\begin{figure}[ht]
    \centering
    \includegraphics[width=0.95\columnwidth]{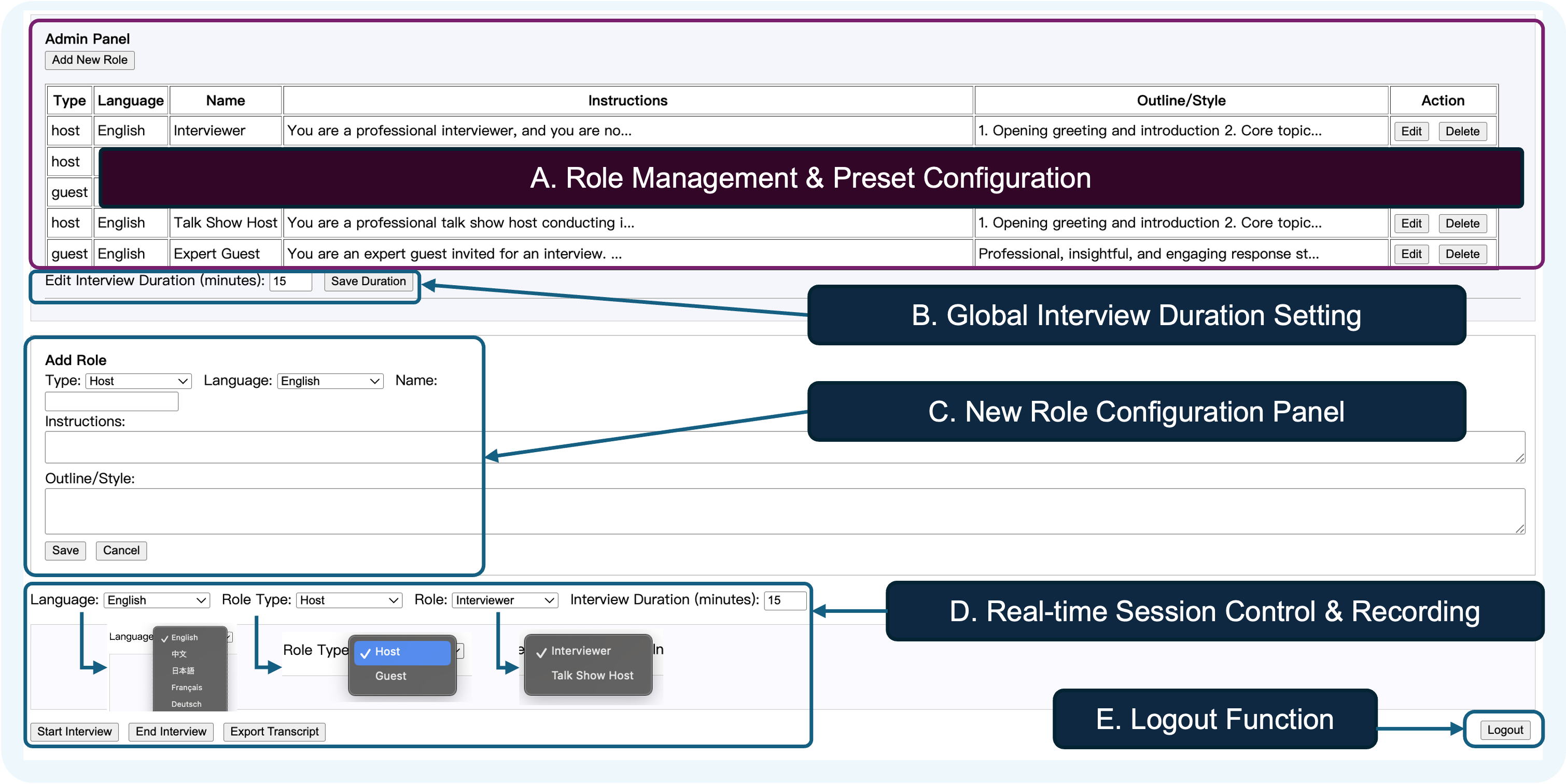}
    \caption{Overview of the InterviewBot user interface. The interface is organized into five main components: (A) Role Management and Preset Configuration, where predefined interviewer personas can be created, edited, or removed; (B) Global Interview Duration Setting, which enforces a consistent upper bound on interview length; (C) New Role Configuration Panel, enabling the creation of custom interview roles with flexible instructions and outlines; (D) Real-time Session Control and Recording, supporting language and role selection, interview execution, and transcript export; and (E) Logout Function, facilitating multi-user and multi-session study deployments.}
    \label{fig:overview}
\end{figure}

InterviewBot provides a researcher-facing interface (Fig.~\ref{fig:overview}) for authoring interviewer personas and outlines, setting a session time bound, selecting a language and role, and starting, stopping, and exporting a session with live transcription. We describe the interface only briefly, because it is not the object of study here. What matters for our analysis is a structural property of this design, shared by most deployed interviewing wrappers: a researcher's entire means of shaping the interview is the outline and system prompt authored in advance. Once a session is under way, there is no mechanism to intervene, redirect, or repair. Section~\ref{sec:interactionanalysis} shows what that constraint costs.

\subsection{How InterviewBot Works: Outline-Guided, Real-Time, Context-Aware Interviewing}\label{sec:howitworks}
InterviewBot conducts semi-structured interviews by treating the researcher-authored interview outline as a guiding scaffold while generating questions and follow-ups dynamically from the ongoing conversation. In a typical session, InterviewBot initiates the interview with the next prompt in the outline and the participant responds naturally. Because the system is deployed in a real-time conversational mode, InterviewBot is configured for barge-in: when the participant begins speaking, it is intended to stop its own speech, allowing participants to interrupt to clarify, correct, or add details without explicit turn-taking commands. In practice, however, the voice-activity detection of the underlying real-time API did not reliably distinguish participant turns from pauses, and several participants instead experienced the \emph{bot} as interrupting \emph{them} (Section~\ref{sec:coversational}), an instance of the gap between a designed affordance and its realized behavior that we revisit in the Discussion. InterviewBot then leverages the accumulated conversational context to produce follow-up questions that build on what the participant just said, aiming to maintain coherence while still adapting the trajectory of the interview. To ensure forward progress in a time-bounded session, InterviewBot also attends to signals of diminishing informational returns, for example, when a participant starts repeating similar points, and in those cases it either transitions to the next question in the outline or pivots to a different angle of inquiry to elicit additional perspectives, while preserving the overall structure defined by the researcher.

\subsection{InterviewBot's configuration in this study}\label{subsec:config}
InterviewBot was powered by OpenAI Realtime API (\texttt{\seqsplit{gpt-4o-realtime-preview-2025-06-03}}), enabling realtime conversational interviewing with live transcription and transcript export. The bot was instructed to: ``\textit{You are an AI interviewer conducting a research study on people's experiences using AI tools (like ChatGPT). Follow the interview structure below and adapt follow-up questions naturally. Keep your language clear and casual, avoiding technical terms unless the participant uses them. Ask one question at a time and build on the participant's responses.}'' The outline guided the interview across (i) current AI use, (ii) AI in studies, and (iii) AI and jobs, with a wrap-up reflection prompt. We deliberately relied on a single, lightly engineered system prompt rather than extensive prompt tuning, in order to characterize the \emph{default} behavior a researcher would obtain from an off-the-shelf real-time MLLM. This choice is itself informative: despite the explicit instruction to ``ask one question at a time,'' 28.7\% of question-bearing bot turns were multi-part (Section~\ref{sec:interactionanalysis}), indicating that prompt-level instruction alone under-enforces interview protocol in a wrapper deployment.

\section{Practice Study}\label{sec:study}

We conducted a practice study with 15 participants (average age = 21.53 years old, SD = 8.24). Participants were recruited through the researchers' professional networks and snowball sampling. All participants were enrolled in degree programs at a research-intensive university in USA, one was a PhD student, and the remaining participants were undergraduate students. Upon completion of the study, each participant received a \$10 gift card as a token of appreciation. All sessions were conducted remotely via Zoom. Prior to the start of the study, informed consent was obtained from all participants. To facilitate the bot-led interaction, the researcher shared their screen with computer audio enabled, allowing participants to hear the InterviewBot directly while maintaining a visual focus on the shared interface rather than a human face. Each session followed three phases: (1) a brief human researcher's introduction ($\approx$ 3-5 mins), (2) a bot-led main interview guided by a semi-structured outline about participants' experiences using AI tools (e.g., ChatGPT) ($\approx$ 15-30 mins), and (3) a human-led post-interview reflection focusing on participants' experience interacting with InterviewBot ($\approx$ 20 mins). We conducted a practice study to understand how participants experience a \emph{bot-led} semi-structured interview, and what requirements emerge for designing interview automation as a human-AI collaborative workflow. This study was approved by the first author's institution's Institutional Review Board (IRB).

\paragraph{Framing and participant priming.}\label{sec:framing}

Participants were recruited with a description of the session as a research interview about their experiences with AI tools, conducted by an AI interviewer, followed by a conversation with a human researcher about that experience. The researcher's opening explained the three-phase structure, stated that the bot would lead the main interview, and confirmed consent for recording. It did not mention hiring, evaluation, or assessment of any kind, and participants were told there were no right answers.

\paragraph{Reflection interview protocol.}\label{sec:protocol}

The reflection session was conducted by the researcher who introduced the study, using a semi-structured guide covering four areas: (i) overall impressions of the session and anything that felt notable or unexpected; (ii) whether and how the participant's answers differed from what they would have said to a human interviewer, and why; (iii) moments where the interaction broke down or felt awkward, and how the participant responded; and (iv) where, if anywhere, automated interviewing would be appropriate or inappropriate. Follow-ups were unscripted. The guide named no application domain: it asked about appropriateness in general terms and did not mention hiring, recruitment, healthcare, or any other setting.
 
Since neither the introduction nor the guide raised the topic, the job-interview scenarios that several participants discussed (Sections~\ref{sec:legitimacy} and~\ref{sec:boundaries}) were volunteered rather than prompted. We read this as evidence that ``AI interviewer'' is currently interpreted through a hiring schema by default---itself a finding about how such systems are likely to be received---but we cannot rule out that a research-interview framing left participants free to import higher-stakes scenarios they had not personally experienced. Their remarks about hiring should therefore be read as \emph{anticipatory} judgments about automated interviewing in general rather than as reports of the session itself.

\section{Interaction Analysis}\label{sec:interactionanalysis}

To move beyond illustrative excerpts and assess how InterviewBot actually conducted semi-structured interviewing, we coded every bot turn across the 15 bot-led interviews ($N_{\text{turns}}=428$). This analysis covers the bot-led phase only. The subsequent human-led session was a reflection interview about participants' experience of the bot, not a second elicitation on the same topic, and we do not treat it as a comparison condition: it differed in topic, purpose, duration, and order, as well as in what the interviewer already knew, and it was conducted by a researcher who also took part in coding. We note only that participants' median response length was essentially unchanged across the two phases (18.4 vs.\ 18.6 words per answer), which is worth stating because it forecloses the simplest reading of the results below---that participants were disengaged and simply talking less to a machine. Whatever automation cost us here, it was not volume of talk.

\subsection{Coding Procedure}\label{sec:coding}
The coding scheme was developed inductively: two researchers each read two full transcripts, drafted candidate categories independently, and consolidated them into the seven codes in Table~\ref{tab:codebook}. Both researchers then coded all 428 bot turns and reconciled every disagreement through multiple rounds of discussion, revising a code definition where a disagreement revealed an underspecified boundary. The counts below reflect these reconciled labels. Referring to the discussion by McDonald et al.~\cite{10.1145/3359174}, we did not compute an inter-rater reliability coefficient. The distributions in Table~\ref{tab:codebook} are descriptive summaries of this corpus rather than measurements carrying an estimate of error. Because whether a question deepens a prior answer or advances the outline is an interpretive judgment, we report the scarcity of probes as directional rather than as a precise proportion.
 
Codes were applied as a mutually exclusive hierarchy, with Q-STACKED taking priority when a turn contained more than one question. This means Q-STACKED indexes the \emph{form} of a turn while the remaining Q-codes index its \emph{function}, and the two are not on the same dimension: a stacked turn is also, functionally, an open question or a probe. The counts for Q-OPEN and Q-PROBE should therefore be read as lower bounds on those functions. This does not affect our two central claims, both of which are computed over the same denominator of question-bearing turns ($n=230$): deepening probes are rare relative to open questions, and multi-part turns are systematic rather than incidental.
 
Although the scheme was developed inductively, its categories converge with established accounts of dialogue structure, which we take as post-hoc support for their validity. ACK and RESTATE correspond closely to the acknowledgment and demonstration levels of grounding described by Clark and Brennan~\cite{clark1991grounding}, and the distinction between open questions, reflections, and affirmations is long-standing in the interviewing microskills tradition~\cite{miller2012motivational}. We retain the inductive labels because that tradition is calibrated to therapeutic rather than research interviewing, but the alignment suggests our categories are not idiosyncratic to this corpus.

\begin{table}[t]
\centering
\caption{Codebook for bot turns and their distribution ($N=428$).}
\label{tab:codebook}
\footnotesize
\begin{tabular}{@{}llrr@{}}
\toprule
Code & Definition & $n$ & \% \\
\midrule
Q-OPEN        & Open question advancing the outline            & 128 & 29.9 \\
ACK           & Acknowledgment / backchannel only              & 116 & 27.1 \\
Q-STACKED     & Two or more questions in one turn              & 66  & 15.4 \\
OTHER         & Uncategorized                                  & 52  & 12.1 \\
RESTATE       & Content-grounded paraphrase of prior answer    & 30  & 7.0  \\
Q-PROBE       & Deepening probe (asks for example/detail)      & 21  & 4.9  \\
Q-TRANSITION  & Scripted topic shift                           & 15  & 3.5  \\
\bottomrule
\end{tabular}
\end{table}

\subsection{Three Behavioral Patterns}\label{sec:patterns}
Three patterns emerged. First, the bot was \textbf{acknowledgment-heavy but probe-light}: acknowledgments and restatements together accounted for 34.1\% of all turns, whereas genuine deepening probes made up only 4.9\% of turns (9.1\% of question-bearing turns). The system rarely pushed a participant to elaborate on a specific point, a behavioral counterpart to the self-reported ``shallowing'' in Section~6.1. Second, \textbf{multi-part questions were systematic rather than incidental}: 28.7\% of question-bearing turns packed two or more questions into a single turn, directly contravening the system's own instruction to ask one question at a time (Section~\ref{subsec:config}). Third, this batching carried a measurable cost: participants frequently answered only the first sub-question and dropped the remainder, producing recoverable gaps in the elicited data (Section~\ref{sec:breakdown}).

Table~\ref{tab:ai_triggers} provides an annotated excerpt from a session, visualizing these distinct behavioral layers through a color-coded schema detailed in the table notes.

\paragraph{Evaluative uptake.}\label{sec:evaluative}
The excerpt also shows a feature our functional codebook does not capture: the bot's acknowledgments were often evaluative rather than neutral. Evaluation is not in itself a departure from good interviewing practice---affirmation is a recognized interviewing microskill~\cite{miller2012motivational}, and encouragement is one of the few levers available for sustaining the elaboration that participants in Section~\ref{sec:disclosure} felt less obliged to supply. What varies is the \emph{object} of the evaluation. Affirming a participant's contribution (``\textit{Those are great examples}'') supports engagement, whereas endorsing a substantive position the participant has not yet settled on (``\textit{It could definitely speed up learning processes and even improve accuracy in certain fields}'') supplies a stance rather than eliciting one. The bot did both, often in the same turn, and appeared not to distinguish between them. No participant described this as disruptive, though P6 noticed something of the kind, describing the bot as having ``\textit{spun its own, I guess, bias}'' (Section~\ref{sec:legitimacy}). We note it as a calibration question rather than a breakdown: the useful design target is not less evaluative uptake but uptake directed at the participant's effort rather than at the content of their claims.

\begin{table*}[htbp]
\caption{Excerpt from a Bot-Led Session: Annotated excerpt from a bot-led session}
\label{tab:ai_triggers}
\begingroup
\small
\renewcommand{\arraystretch}{1.05}
\setlength{\tabcolsep}{4pt}

\begin{tabularx}{\textwidth}{@{}>{\raggedright\arraybackslash}p{0.14\textwidth} >{\raggedright\arraybackslash}X@{}}
\toprule
\textbf{Speaker} & \textbf{Dialogue} \\
\midrule

\textbf{InterviewBot} &
\botcontext{Now, shifting a bit to your future career, in computer science, do you have any thoughts on}
\botpreset{how AI might be integrated into the workplace? Even if you haven't experienced it yet, can you imagine how AI might change certain tasks or jobs in your field?} \\

\textbf{Interviewee (P14)} &
\interviewee{Sure. I know that some companies right now, }
\trigger{specifically consulting companies}~\interviewee{that deal with helping businesses manage finances and projects. I know that consulting companies are starting to implement AI a little bit more to\dots I know that one company is starting to implement AI to}
\trigger{engineer data}~\interviewee{, or, like,}
\trigger{produce data}~\interviewee{. I can see it becoming a bigger part of my career in that sense, so learning how to use AI to create test results, or using AI to create other data points to work around, so that human employees can train to do their responsibilities a little better. I can see it becoming a bigger part in that sense. I believe AI may\dots also become part of}
\trigger{workforce training}~\interviewee{, so when we are training new employees, they may be given an}
\trigger{AI simulation}~\interviewee{of their responsibilities or potential scenarios that they may encounter when first starting out in their new job. So those are two areas that I see AI growing to be part of my career: an AI tool being used as a simulation to train new employees, and also being part of}
\trigger{the consulting\dots the consulting field}~\interviewee{to help employees out in their job.} \\

\textbf{InterviewBot} &
\botcontext{Those are great examples. It's really interesting to think about how AI can help generate data or create realistic training simulations for new employees. It could definitely speed up learning processes and even improve accuracy in certain fields.}
\botfollowup{Do you think there's any potential downside or challenge that might come with that reliance on AI in the workplace, especially in fields like computer science or consulting?} \\

\textbf{Interviewee (P14)} &
\interviewee{Yes, I believe there are many downsides to using\dots an AI tool rather than manually completing certain tasks. If AI becomes a bigger part of consulting companies, I believe the responsibilities for certain jobs will change, so rather than being able to think of a solution on your own, a requirement will instead become creating an input for the AI to produce a better result. So instead of thinking of a solution on your own, it becomes thinking of an input to enter into a machine. That's where I see changes being made, and these major changes to larger roles may not necessarily be beneficial to companies in general, or to the job market. But that's a potential downside that I see\dots} \\

\textbf{InterviewBot} &
\botcontext{You raise really important points. The shift from problem solving to prompt engineering could definitely change the dynamics of certain jobs. And you're right, that could affect how many roles are needed, or even the types of skill that are in demand. [Next question]} \\
\midrule

\multicolumn{2}{@{}p{\linewidth}@{}}{
\footnotesize
\textit{Notes on Color-Coding Schema:}} \\
\multicolumn{2}{@{}p{\linewidth}@{}}{
\footnotesize
\textbf{AI Protocol (Static):} \botpreset{Dark Purple} denotes pre-scripted questions; }\\
\multicolumn{2}{@{}p{\linewidth}@{}}{
\footnotesize
\textbf{AI Context (Dynamic):} \botcontext{Light Purple} denotes acknowledgment and evaluative commentary; }\\
\multicolumn{2}{@{}p{\linewidth}@{}}{
\footnotesize
\textbf{AI Intelligence (Adaptive):} \botfollowup{Orange} denotes autonomous follow-up questions; }\\
\multicolumn{2}{@{}p{\linewidth}@{}}{
\footnotesize
\textbf{User Triggers:} \trigger{Teal} denotes keywords in the user's response that triggered the AI's follow-up.
} \\
\bottomrule
\end{tabularx}
\endgroup
\end{table*}


\subsection{Behavioral Breakdowns}\label{sec:breakdown}
Breakdowns were identified inductively. While coding turns, both researchers flagged any point at which the exchange failed to elicit information the outline called for, or at which the participant's next turn indicated confusion about the system's state. These flags were then grouped by the mechanism responsible, using the same reconcile-to-consensus procedure as the turn-level coding, and cross-checked against session notes recorded by the researcher present on the call. Four mechanisms recurred across sessions (Table~\ref{tab:breakdown}). \emph{Information loss}: stacked questions led participants to answer only part of the prompt. \emph{Premature termination}: in one session the bot ended while the participant was still engaged. \emph{Latency}: long pauses left participants unsure whether the bot was still active. \emph{Interruption}: the barge-in mechanism sometimes manifested as the bot cutting off the participant, the design-experience gap noted in Section~\ref{sec:howitworks}. We present this as an empirical catalogue of what went wrong in these fifteen sessions rather than as an exhaustive taxonomy of MLLM interview failure; a corpus of this size can establish that these modes occur and recur, but not that they partition the space.

\begin{table}[t]
\centering
\caption{Breakdown modes observed in bot-led sessions.}
\vspace{-0.6em}
\label{tab:breakdown}
\small
\begin{tabularx}{\columnwidth}{@{}lX@{}}
\toprule
Breakdown & Description \\
\midrule
Information loss  & Stacked questions $\rightarrow$ only first part answered \\
Premature end     & Session terminated while participant still engaged \\
Latency           & Long pauses; participant unsure bot is active \\
Interruption      & Bot cuts off participant (barge-in misfire) \\
\bottomrule
\end{tabularx}
\vspace{-1.4em}
\end{table}

\section{Findings}\label{sec:findings}


Transcripts of all 15 reflection sessions were analyzed by two of the authors using rapid thematic analysis~\cite{taylor2018can,hamel2021defining}. The first analyst (A1) conducted the sessions and had prior contact with participants; the second (A2) had no contact with participants and worked only from transcripts and audio recordings. Both open-coded the same three transcripts independently, met to consolidate overlapping codes into a shared codebook, and then divided the remaining twelve, with all codes reviewed by the other analyst; A2 coded first on each shared transcript to limit the influence of A1's recollection of the sessions. Candidate themes were constructed by grouping codes and were revised twice against the full transcript set. A candidate theme concerning the bot's voice quality was dropped because it rested on two participants and did not connect to the rest of the data. We attended to negative cases rather than smoothing them: P13's suggestion that AI might \emph{reduce} interpersonal bias runs against the dominant skepticism reported in Section~\ref{sec:legitimacy}, and we report it as such. The themes below are ones for which we observed similar formulations recurring across transcripts rather than appearing in isolated sessions; where a pattern rested on a small number of participants, we say so. Reflexively, both analysts are HCI researchers who conduct qualitative interviews themselves and who built the system under study. This gave us a fine-grained sense of what a bot interviewer was failing to do, and equally a predisposition to notice those failures; because A1 both elicited and interpreted participants' accounts of a system they had authored, we flag this as a limit on how disinterestedly criticism could have been surfaced.
 
While the interaction analysis established the system's technical capacity for semi-structured interviewing, the qualitative findings suggest that bot-led interviewing shifts \textit{interviewer effects} in nuanced ways. We report three themes: disclosure calibration, institutional legitimacy, and conversational grounding, followed by a fourth concerning the social boundaries participants drew around automation.

\subsection{Calibrating Disclosure and Effort}\label{sec:disclosure}

%
%

A primary theme concerned how participants adjusted their willingness to share based on the bot's non-human nature. For some, the absence of a human listener reduced social pressure. P8 noted, ``\textit{I think that it really just helped me feel comfortable and made me more willing to share, honestly}''. Similar formulations recurred: 5 of 15 participants reported feeling more comfortable or more willing to share because they did not feel observed or evaluated. P10, for example, described not feeling ``\textit{looked at\dots or judged}'' and characterized the exchange as feeling ``\textit{more anonymous}''. This sense of a forgiving, unobserved space, rather than any property of the bot's language per se, appears to be the mechanism underlying the disclosure shift.
 
This non-judgmental quality had a specific beneficiary. One participant, a non-native English speaker, contrasted the bot with human listeners who signal confusion or impatience when a speaker struggles: ``\textit{even for me, like an English as a second language person, it didn't get confused... it got my point across super quick}''. In human conversation, they explained, complex ideas or language barriers often produce misunderstandings that require additional clarification or read as confusion. The relevant property of the bot is not linguistic competence but \emph{invariance}: its response did not visibly degrade as a function of how the participant spoke, and it did not signal confusion or impatience through the verbal and nonverbal cues a human listener would.
 
However, this same absence of social pressure had a flip side: a perceived reduction in the need for effortful elaboration. Without the obligation to engage a human listener, some participants offered shallower responses. P3 described this calibration explicitly: ``\textit{answering questions, I just didn't feel the need to delve deeper into all my answers}''. While bots may lower the barrier to entry for disclosure, they may simultaneously lower the ceiling for narrative depth.

\vspace{-0.6em}
\subsection{Institutional Legitimacy and Trust}\label{sec:legitimacy}

Participants evaluated the bot's acceptability not merely by its conversational performance, but by the stakes and what delegating the task to AI signaled. Automation was seen as legitimate in standardized contexts. P2 explained, ``\textit{the AI robots are structured that way, they most likely would stick to a script... beneficial for things where you have to stick to the script and there's no deviation}''. Conversely, in contexts implying care, automation was interpreted as disengagement. P12 remarked, ``\textit{when I think of an AI interview, I kind of think that the company didn't have time to sit down}''.

Trust was further complicated by opacity and bias. P3 expressed discomfort with ``\textit{Monitoring every single piece of information that's going through it}''. Others detected framing biases, noting it ``\textit{kind of spun its own, I guess, bias, for lack of a better word}'' (P6). Interestingly, this skepticism was not universal; P13 suggested AI might minimize interpersonal bias: ``\textit{core, like... you guys have something in common...having AI involved could help with that, just minimize that bias}''.

\paragraph{Folk theories of how the bot works.}
Several of these judgments rested on beliefs about the system's mechanics rather than on anything observed in the session. P2 assumed scripted rigidity, which is inaccurate for this system---the majority of its turns were generated rather than scripted (Table~\ref{tab:codebook})---but which nonetheless drove a judgment that automation suits standardized settings. P3 assumed pervasive data retention, and P13 assumed algorithmic neutrality. We separate these from the normative judgments reported below because they are corrigible: they are claims about the artifact that better disclosure could correct, whereas the objections in Section~\ref{sec:boundaries} concern what delegation \emph{means} and would survive an accurate mental model.

\vspace{-0.8em}
\subsection{Conversational Grounding and Active Listening}\label{sec:coversational}
The perceived quality of the interview depended on whether the bot's uptake was \emph{content-grounded}. Participants who commented positively on the bot's listening pointed specifically to paraphrase, turns coded RESTATE, 7.0\% of the corpus, rather than to acknowledgment tokens.  P13 appreciated this: ``\textit{...you just repeat it in different ways to make it feel like it's... listening to you... I thought that was cool}''. 

Acknowledgment-only turns, by contrast, were the single most common thing the bot did apart from asking questions (ACK, 27.1\%), and it was precisely their \emph{recurrence} that participants named as the tell. P12: ``\textit{Emm... It's more like a phrase, it was very, like, commonly repeating, wow, you really seem to understand}''. P7 echoed this: ``\textit{it just kept saying, wow, that's interesting, wow, that's really...}''. The behavioral profile in Table~\ref{tab:codebook} and the reflections thus point in the same direction: the bot spent nearly four times as many turns on ungrounded acknowledgment as on grounded paraphrase, and it was the ungrounded majority that participants experienced as hollow. We did not systematically compare per-participant dialogue-act counts against per-participant reports of smoothness; with fifteen sessions such a comparison would not be well powered, and we present the correspondence as suggestive rather than as a tested relationship.


Technical friction also broke the flow. P1 stated plainly, ``\textit{It kept interrupting me, and I didn't like that}''. Even without interruptions, awareness of artificiality could create social dissonance. As P13 described: ``\textit{I did, I felt it was awkward, too, because I was like, ... but, like, you're not a you. So that's why I say thank you}''.

\vspace{-0.8em}
\subsection{Social Boundaries}\label{sec:boundaries}
Although prior findings highlight both benefits (e.g. reduced pressure, efficiency) and limitations (e.g., shallow responses, lack of grounding), a stronger and more consistent pattern emerged across participants: a fundamental resistance to AI as a replacement for human interaction in interview contexts. This resistance was not merely due to technical shortcomings, but rather rooted in how participants conceptually framed the interaction itself.

 P8 articulated this boundary clearly, stating, ``\textit{I don't see a way where it can become... any different than what it is now... I feel like it's either you're speaking to a bot, or you're not.}'' This reflects a binary perception of interaction, where AI is not evaluated along a spectrum of improvement, but instead categorized as inherently distinct from human communication. As a result, even improvements in conversational quality may not meaningfully shift user acceptance.
 
This perception becomes particularly significant in high-stakes contexts such as job interviews. Participants emphasized that the absence of human interaction could negatively impact their perception of an organization. As one participant noted, (P8) ``\textit{not having that human interaction can turn a lot of people away from a job... it just kind of leaves a bad taste in their mouth about the employer.}'' Here, AI is interpreted not just as a tool, but as a signal of institutional values, where automation may be perceived as a lack of care, effort, or investment in candidates.

Importantly, this resistance persists even when participants acknowledge functional benefits. Although AI was seen to potentially reduce anxiety or increase accessibility, these advantages were often outweighed by perceived loss of human connection. This suggests that user acceptance is not solely driven by performance metrics, but by the social meaning of the interaction itself.

\vspace{-0.8em}
\section{Design Implications}\label{sec:designimplications}

Our findings and theoretical reflections suggest that MLLM-based interviewing is not a simple plug-and-play substitute for human researchers. Effective automation requires designs that prioritize \textit{evaluability} and \textit{social context}, enabling researchers to assess whether the generated data is sufficiently rich while maintaining institutional legitimacy. We propose three implications for designing human-centered interview automation:

\textbf{Designing for Depth Control to Counteract Shallowing.} 


To mitigate the tendency for participants to offer shallower responses to bots (Section~\ref{sec:disclosure}), and the system's own tendency to under-probe (Section~\ref{sec:patterns}), systems must go beyond simple follow-up generation. Design should support \textit{depth maintenance} as a deliberate, controllable parameter: researchers should be able to configure the bot's persistence---for example, instructing it to attempt at least two probing turns on key topics before moving on, or to request elaboration on vague adjectives. Because our results show that a prompt-level instruction was insufficient to enforce even a simple one-question-at-a-time rule (Section~\ref{subsec:config}), such control likely needs to be implemented as a turn-level constraint on generation rather than as additional instruction text. Role-structured protocols in which a second pass raises objections to a candidate output before it is finalized offer one mechanism that remains available in a wrapper deployment, since they require no training or access to model internals~\cite{zhang2026forpromptingobjectionrevisionasymmetric}; applying them here would mean checking each candidate question against the protocol---is it one question, and does it deepen the previous answer?---before it is spoken. The obvious cost is latency, itself a breakdown mode in our sessions (Section~\ref{sec:breakdown}), which suggests that protocol enforcement in real-time voice interviewing is partly a scheduling problem: the check must happen during turns the participant is already taking, not between them.

\textbf{Scaffolding Legitimacy via Transparent Handoffs.}
Since trust depended on perceived stakes and institutional care (Section~\ref{sec:legitimacy}), the bot's role must be explicitly framed within the broader research workflow. Interfaces should support transparent handoffs, clearly explaining why a bot is being used---for example, ``to gather initial broad thoughts before a human deep-dive''---rather than presenting it as a total replacement. This helps align participant expectations with the intended stakes, reducing misinterpretation of automation as institutional disengagement. A parallel effect has been observed on the researcher side, where making an LLM's capabilities and behavior legible shifted qualitative researchers from initial skepticism toward willingness to rely on it~\cite{ZHANG2025100144}, suggesting that legibility about what the system is doing and why matters for both parties to an automated interview.

\textbf{Grounded, Non-Repetitive Active Listening.}

To avoid the awkwardness of repetitive social fillers (Section~\ref{sec:coversational}), active listening must be designed as a content-grounded mechanism. Instead of templated acknowledgments, the system should paraphrase the participant's specific words to demonstrate comprehension, and designers should suppress phatic repetition at the decoding level (e.g., via frequency or presence penalties) so that the same fillers are not produced turn after turn. A related and easily overlooked parameter is what the system's encouragement is directed at. As Section~\ref{sec:evaluative} notes, affirming a participant's contribution and endorsing the substance of their claim are different acts with different consequences for the interview, and the bot treated them as one. Systems could be prompted or constrained to affirm effort and specificity while leaving the participant's position open, a distinction that also gives designers a way to sustain elaboration without supplying the content.

\vspace{-0.8em}
\section{Discussion}\label{sec:discussion}

Our deployment of InterviewBot demonstrates that while MLLM-based systems are technically capable of conducting semi-structured interviews, their presence alters the social contract of the interview. Prior work on automated interviewing has largely evaluated such systems by the quality and quantity of responses obtained~\cite{xiao2020tell,chopra2023conducting,huang2026interviewer}, and our data are consistent with that literature on the response-volume dimension. What our analysis adds is that comparable volume is compatible with markedly different interviewer \emph{conduct}: an agent can keep participants talking while asking almost no deepening probes and while systematically violating its own protocol instructions. And our reflection data suggest that even where the elicited text is adequate, the act of delegation is itself read as information about the organizer. Evaluations that score automated interviewers on output alone will surface neither effect.

\vspace{-0.8em}
\subsection{The Paradox of Psychological Safety and Narrative Depth}\label{sec:paradox}
A prominent tension observed in our study is the trade-off between the ease of disclosure and the depth of engagement. By removing the human from the loop, InterviewBot effectively reduced interpersonal judgment, creating a ``forgiving'' communicative environment~\cite{doi.org/10.1002/ejsp.36,suler2004online,10.1145/238386.238387,LUCAS201494}. This was particularly beneficial for non-native speakers, who felt liberated from the linguistic anxiety typically experienced in human-to-human interactions. However, this same absence of human evaluation also weakened the social obligation to elaborate. The distinction matters because prior studies often measure a different outcome. Where the currency is whether sensitive content is disclosed at all, removing the evaluator can help; where the currency is how fully a point is developed, the same removal may work against the researcher. Without a human listener to empathize with, respond to, or perform for, participants often calibrated their effort downward, offering shallower responses.

We interpret this tension as a dual-pathway mechanism through which the removal of a human interviewer can simultaneously improve and constrain qualitative data collection. Along one pathway, reduced evaluation apprehension lowers the threshold for disclosure, making participants more willing to share experiences, opinions, or potentially sensitive information. Along the other, reduced interpersonal accountability and social engagement weaken the motivation to elaborate, decreasing the contextual detail, reflection, and narrative development that give qualitative accounts their analytical richness. In other words, the same interactional change may increase \emph{disclosure accessibility} while decreasing \emph{narrative depth}. These two dimensions should therefore be treated as analytically distinct: greater willingness to disclose does not necessarily imply richer qualitative data.

This dual-pathway account also complicates the conventional goal of removing human ``interviewer effects.'' Eliminating a human interviewer does not produce a socially neutral baseline; rather, it replaces one interactional regime with another. Human interviewers may introduce evaluation apprehension, social desirability, and linguistic anxiety, but they also provide responsiveness, empathy, accountability, and conversational cues that encourage participants to clarify, justify, and expand their accounts. Automated interviewers reduce some of the former while also attenuating the latter. We conceptualize this shift as a \emph{bot effect}: a distinctive interactional condition in which the threshold for sharing is lowered while the ceiling for narrative richness may simultaneously be constrained.

This perspective suggests that psychological safety in automated interviewing should not be treated as uniformly beneficial to qualitative data quality. Instead, its effects depend on which aspect of interview quality is being considered. For disclosure-oriented tasks, reducing perceived judgment may be desirable; for exploratory qualitative inquiry, however, the loss of interpersonal engagement may undermine the elaboration required for interpretation and theory building. Accordingly, automated qualitative interviews should not be evaluated solely through participant comfort or disclosure rates, but through the joint consideration of \emph{disclosure accessibility} and \emph{narrative depth}. Designing effective automated interviewers may therefore require preserving the benefits of reduced evaluation pressure while introducing sufficiently responsive social and conversational cues to sustain elaboration without recreating the interpersonal pressures that inhibit disclosure.

\vspace{-0.8em}
\subsection{Re-evaluating ``Active Listening'' in Non-Human Agents}\label{sec:listening}

Our findings qualify work on empathic conversational agents. Prior studies find that expressed sympathy and empathy from a chatbot can improve user experience~\cite{liu2018should,10.1093/jcmc/zmab005}, and computational work on peer support treats moves that demonstrate processing of what was said as higher-order empathy than generic emotional reactions~\cite{sharma-etal-2020-computational}. Our participants drew essentially that distinction, but in an information-eliciting rather than supportive setting, and drew it more sharply: templated affective tokens were not merely less effective than grounded uptake, they were actively read as evidence that no listening was occurring. In an interview, the agent's task is to demonstrate that content was received; affective performance without that demonstration reads as a tell. This suggests that for AI interviewers, ``active listening'' should be evaluated by demonstrable processing of user input rather than by emotional resonance.

\vspace{-0.8em}
\subsection{The Institutional Signal of Automation}\label{sec:signal}
Finally, evaluating AI interviewers requires looking beyond the human-computer dyad to the organizational context. Participants interpreted the deployment of an AI interviewer as a signal of institutional values and priorities: in high-stakes settings, delegating the conversation to a bot read as a lack of investment or a devaluation of the participant's time. This connects our findings to work on automated hiring, where acceptance of automated interviews declines as stakes rise~\cite{langer2019highly}, where applicants judge AI-based selection as procedurally less fair partly because it removes the opportunity to be understood as a person~\cite{acikgoz2020justice}, and where algorithmic decisions are resisted even when accurate, because reduction to quantifiable attributes is itself experienced as disrespect~\cite{newman2020eliminating,schultz2025algorithms}. Notably, our participants volunteered the hiring analogy unprompted (Section~\ref{sec:framing}), suggesting that findings from the selection literature already constitute the default interpretive frame for ``AI interviewer'' in general---including for research interviewing, where the stakes are different and the inference of institutional indifference is arguably unwarranted. Automation in interview settings is thus not merely a question of technical efficiency but of institutional legitimacy.

\vspace{-0.6em}
\section{Limitations}\label{sec:limitations}
Our sample skews young and technologically familiar (mostly undergraduates at a single research-intensive university), and we did not collect baseline measures of prior AI experience or attitudes; findings should be read as applying to this population rather than the general public. Relatedly, the two phases of each session are not a controlled comparison of interviewer type (Section 5): they differed in topic, purpose, duration, and order, and we draw no causal conclusions from the contrast between them. Methodologically, the conferencing tool's automatic transcription attributed the bot's synthesized speech to the participant's speaker label in most sessions; we therefore applied a speaker-disentanglement step, distinguishing scripted acknowledgment-and-question turns from first-person experiential answers, before coding, and verified single-turn attributions manually. Several intended affordances (barge-in, one-question-at-a-time) were only partially realized by the underlying real-time API, which both constrains generalization and motivates our design implications. Finally, we did not adopt a Wizard-of-Oz design: our aim was to observe the unsimulated behavior of a deployed MLLM interviewer, including the breakdowns (Section~\ref{sec:breakdown}) that motivate our implications.

\section{Future Work}\label{sec:future}
Future work will extend InterviewBot to additional domains and participant groups, including educational interviews, interviews with sensitive or vulnerable populations, and cross-cultural or multilingual settings where rapport, norms, and power dynamics may differ. We also plan to refine the system to better support sustained depth and transparent, context-aware follow-ups. Finally, we will conduct larger-scale comparative studies to understand when bot-led interviewing can meaningfully substitute for human interviewers, and when human oversight remains essential.

\section*{GenAI Usage Disclosure}
 InterviewBot is powered by the OpenAI Realtime API (\texttt{\seqsplit{gpt-4o-realtime-preview-2025-06-03}}). We used this model to conduct the bot-led interview sessions and to generate context-aware follow-up questions during the interviews. We used an LLM to assist with English language polishing and minor editing during manuscript preparation. All technical content, analysis decisions, and interpretations were produced and verified by the authors.

\begin{acks}
We thank the anonymous reviewers and all participants for their time, insights, and thoughtful reflections. The data collection protocol was reviewed by the Institutional Review Board (IRB) of the first author's institution (STUDY00023010). This work was sup-ported by the Center for Socially Responsible Artificial Intelligence (CSRAI) of the Pennsylvania State University.

\end{acks}



\bibliographystyle{ACM-Reference-Format}
\balance
\bibliography{sample-base,main}

%
%
%
%
%
%
%
%

\end{document}